%% file: main.tex
\documentclass[aps,reprint,superscriptaddress,nofootinbib,amsmath,amssymb]{revtex4-1}

\usepackage{array}
\usepackage{makecell}
\usepackage{tabularx}
\usepackage{placeins}
\usepackage{multirow}
\usepackage{booktabs}
\usepackage{gensymb}
\usepackage{textcomp}
\usepackage{float}
\usepackage{times}
\usepackage{graphicx}
\usepackage{dcolumn}
\usepackage{bm}
\usepackage{lipsum}  
\usepackage{xcolor}
\usepackage{comment}
\usepackage{lineno}
\usepackage[colorlinks=true,linkcolor=blue,citecolor=blue,urlcolor=blue]{hyperref}

\newcommand{\germania}{GeO$_{2}$}
\newcommand{\silica}{SiO$_{2}$}
\newcommand{\tantala}{Ta$_{2}$O$_{5}$}
\newcommand{\titania}{TiO$_{2}$}
\newcommand{\tdg}{TiO$_2$-doped GeO$_2$}

\usepackage{xspace}

\begin{document}

\preprint{APS/123-QED}

\title{Intrinsic Defects in Amorphous Optical Coatings of TiO$_2$-doped GeO$_2$ for Gravitational-wave Detectors}

\author{K.\,Prasai}
\email{kprasai@kennesaw.edu}
\affiliation{Department of Physics, Kennesaw State University, Marietta, GA 30060, USA}

\author{K.\,Lee}
\affiliation{Department of Physics, Sungkyunkwan University, Seoul 03063, Republic of Korea}

\author{R.\,Bassiri}
\affiliation{E. L. Ginzton Laboratory, Stanford University, Stanford, California 94305, USA}

\author{A.\,Davenport}
\affiliation{Department of Electrical and Computer Engineering, Colorado State University, Fort Collins, Colorado 80523, USA}

\author{D.\,A.\,Drabold}
\affiliation{Physics and Astronomy Department, Ohio University, Athens 45701, USA}

\author{M.\,M.\,Fejer}
\affiliation{E. L. Ginzton Laboratory, Stanford University, Stanford, California 94305, USA}

\author{A.\,Markosyan}
\affiliation{E. L. Ginzton Laboratory, Stanford University, Stanford, California 94305, USA}

\author{C.\,S.\,Menoni}
\affiliation{Department of Electrical and Computer Engineering, Colorado State University, Fort Collins, Colorado 80523, USA}

\author{S.\,Tait}
\affiliation{LIGO Laboratory, California Institute of Technology, Pasadena, California 91125, USA}

\date{\today}

\begin{abstract}
\noindent The increased laser power of future gravitational-wave detectors will require mirror coatings with optical absorption below 0.1 ppm per mirror. TiO$_2$-doped GeO$_2$, currently the best high-index material for reducing room-temperature coating thermal noise, still exhibits ppm-level absorption even after extrinsic contamination is minimized. Using {\it ab-initio} simulations and absorption measurements, we identify oxygen-deficient Ti-rich environments as the origin of this residual absorption. We find that ordinary structural disorder in the amorphous network can localize electronic states but does not produce defects capable of absorbing 1064-nm light. In contrast, oxygen vacancies in compact Ti-rich environments create localized Ti$^{3+}$--Ti$^{3+}$-like polaron-pair or mixed Ti-polaron states states with transitions near 1064 nm. Photothermal measurements show increased absorption after dry/inert annealing, supporting the formation of these reduction-sensitive defects. These results show that the residual absorption is not an intrinsic limitation of TiO$_2$-doped GeO$_2$, but a process-dependent defect that may be mitigated through control of oxygen stoichiometry during deposition and annealing.
\end{abstract}

\pacs{Valid PACS appear here}

\maketitle
The sensitivity of interferometric gravitational-wave detectors depends critically on the thermal-noise and optical performance of their mirror coatings \cite{steinlechner2018development}. These coatings must provide extremely high reflectivity while maintaining low mechanical dissipation and low optical absorption. In Advanced LIGO and Advanced Virgo, which operate at room temperature with a laser wavelength of 1064 nm, the high-reflectivity mirrors are formed from alternating layers of ion-beam-deposited \titania-doped \tantala\ and \silica\ in a Bragg structure \cite{abbott2016observation,collaboration2015advanced,acernese2015advanced,harry2006titania}. In the part of the detection band where the detectors are most sensitive, approximately 50--300 Hz, the sensitivity is limited by coating Brownian thermal noise \cite{martynov2016sensitivity,buikema2020sensitivity}. Reducing this noise while preserving or improving excellent optical performance is therefore a central requirement for upgrades of current and future detectors.

This need has motivated the search for alternative high-index amorphous materials with lower mechanical loss than \titania-doped \tantala. Among the candidates, amorphous \tdg, paired with \silica\ as the low-index layer, has emerged as a leading option for room-temperature gravitational-wave detectors. Previous work showed that GeO$_2$-based glasses provide a suitable low-mechanical-loss high-index host, and that Ti incorporation can raise the refractive index while preserving sufficiently low mechanical loss for Advanced LIGO's next upgrade (called Advanced LIGO+ or A+) \cite{vajente2021low}. Guided by this understanding,  \tdg/\silica\ coatings were down-selected for the Advanced LIGO+ coating Pathfinder process and were shown to achieve substantially reduced coating thermal noise relative to the current LIGO coatings \cite{AplusWP,vajente2021low}. At the same time, however, optical absorption remains a major concern for A+ upgrade as even when extrinsic contamination is reduced, the measured absorption of \tdg/\silica\ high-reflectivity stacks at 1064 nm remains at the ppm level, and depends sensitively on deposition and annealing conditions \cite{CEcoatingsWP}. Previous PCI measurements at $\lambda=1064$ nm found that a \tdg\ film with Ti/$(\mathrm{Ti}+\mathrm{Ge})\simeq 0.44$ and refractive index $n=1.88$ had an absorption of $2.3\pm0.1$ ppm when normalized to a quarter-wave optical thickness of 141 nm after annealing at $600\,^\circ$C and a multilayer \tdg/SiO$_2$ stack fabricated from the same material system showed an absorption of about 3.1 ppm after annealing \cite{vajente2021low}. These values show that \tdg\ is already a low-absorption material on an absolute scale, but also that the remaining absorption is technologically important. For future detectors, as the detector circulating power continues to increase, coating absorption will place an increasingly stringent constraints on detector performance as the absorbed laser power leads to heating, thermal lensing, and distortion of the optical mode. Cosmic Explorer requirements motivate a further reduction toward $\sim 0.1$ ppm per mirror \cite{CEcoatingsWP}. Understanding the origin of this residual absorption is therefore essential. 

Because 1064 nm photons are far below the fundamental band gap of these oxides, the absorption cannot arise from ordinary band-to-band transitions or from Urbach tails. In amorphous oxides, it is expected to originate from localized states within the gap or near the band edges, often related to oxygen defects, nonstoichiometric local environments, hydroxyl-related species, transition-metal impurities, and Ti-related redox or coordination motifs \cite{o1983theory,humbach1996analysis,beales1976special}. In \tdg, this problem is especially subtle because the material does not contain a single well-defined local structure. Prior studies using grazing-incidence pair distribution function (GIPDF) measurements, Raman spectroscopy, x-ray photoelectron spectroscopy (XPS), and atomistic modeling have already shown that Ti can occupy multiple local bonding environments and that the population of these motifs evolves with composition and annealing \cite{bhowmick2024unveiling, zhang2026network, prasai-submitted}. It is therefore natural to ask whether some of these local environments also act as intrinsic absorption centers. 

Structural models of amorphous \tdg\ were generated using a melt--quench \textit{ab-initio} molecular dynamics (AIMD) workflow \cite{drabold2009topics}. We began from AIMD-generated model of amorphous \germania, following the procedure described in Ref.~\cite{prasai2023glass}. The model contained 360 atoms. To obtain \tdg, 40\% of the Ge atoms in these parent \germania\ models were replaced at random by Ti atoms. This Ti concentration was chosen to remain close to the composition identified as optimal for next-generation LIGO coating applications \cite{vajente2021low}. The simulation cell was then rescaled to give an initial density of 3.65 g/cm$^{3}$, consistent with the experimental density reported for 40\% \tdg\ thin films (3.73 and 3.62 g/cm$^{3}$ for as-deposited and annealed films respectively, Ref. \cite{prasai-submitted}).

The subsequent preparation was carried out in two stages. Stage 1 was used to optimize the density in the following way. Each model was equilibrated at 800 K for 1 ps, cooled to 300 K over 2 ps, and then equilibrated at 300 K for 5 ps. Five snapshots were taken from the final 300 K trajectory and each was subjected to simultaneous ionic and cell-volume relaxation using the conjugate-gradient algorithm. The relaxed densities of these five snapshots were averaged, yielding 3.77 g/cm$^{3}$ with standard deviation of 0.005 g/cm$^{3}$. The final configuration from the 300 K AIMD trajectory was then rescaled to this average density and used as the starting point Stage 2.

In Stage 2, melt--quench cycles were carried out \emph{two times}. This repeated melting and quenching was carried out to reduce memory of the initial substitutional starting structure and to allow the mixed Ge--Ti--O network to reorganize more fully, which is important in melt--quench modeling of multicomponent amorphous oxides \cite{zhao2005structural, yan2012multiscale}.

In the first melt--quench cycle, the model was equilibrated at 2000 K for 10 ps, cooled to 300 K over 10 ps, and equilibrated at 300 K for another 10 ps. In the second melt-quench cycle, the model was equilibrated once more, now at 2500 K for 50 ps. During the final 10000 steps of this trajectory, ten snapshots were extracted at intervals of 1000 steps. Each snapshot was then independently quenched to 300 K over 25 ps using a different random initialization of atomic velocities, followed by equilibration at 300 K for 10 ps. Finally, each quenched configuration was relaxed to its local energy minimum using the conjugate-gradient algorithm. The resulting set of relaxed structures was used for all subsequent analyses. This ensemble-based procedure was adopted because a single quenched configuration is generally not sufficient to represent the structural variability of an amorphous network.

\begin{figure}[t]
	\centering
		\includegraphics[width=\linewidth]{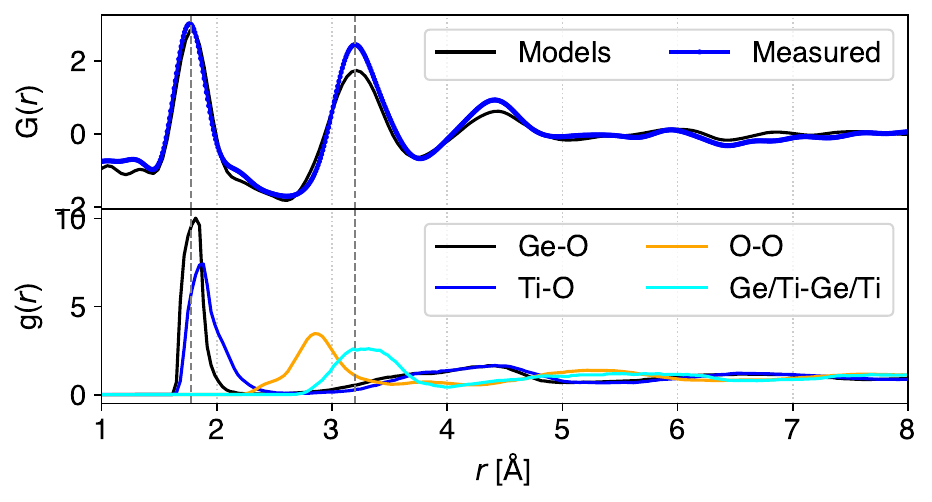}
		\caption{Pair distribution functions (PDF). Top: X-ray total PDF computed from models in this work and x-ray GIPDF measured in Ref \cite{prasai-submitted}. Bottom: Partial PDF computed from models in this work. To reduce noise, all computed PDF are smoothed using running window average of size 6. The positions of first two major peaks in x-ray PDF of models are marked at 1.78 \AA~and 3.20 \AA~on both plots.}  
		\label{fig:PDF}
\end{figure} 

All AIMD and electronic-structure calculations were performed using the Vienna \textit{ab initio} simulation package (VASP) \cite{kresse1993ab,kresse1996efficient}. Electron-ion interactions were treated using the projector augmented-wave (PAW) method \cite{blochl1994projector}. The electron exchange and correlation were estimated using (a) the Perdew--Burke--Ernzerhof (PBE) generalized-gradient approximation \cite{perdew1996generalized} for all MD simulations, (b) the PBE+\(U\) approach \cite{dudarev1998electron} for O-vacancy relaxations in which \(U_{\rm eff}=U-J=4.5\) eV is applied to Ti \(3d\) states, partially correcting the PBE self-interaction error on localized Ti \(3d\) states, helping to stabilize vacancy-induced electron localization, and (c) HSE06 approach \cite{krukau2006influence} for electronic structure calculations on final structure models. The HSE06 is a screened hybrid functional built on the PBE in which a fraction of short-range PBE exchange is replaced by Hartree--Fock exact exchange, generally improving band gaps and defect-state energetics relative to PBE while remaining an approximate density-functional treatment. The HSE06 calculations employed the standard exact-exchange mixing fraction of 25\% and a range-separation parameter of 0.20 \AA$^{-1}$. Valence-electron orbitals were expanded in a plane-wave basis with a kinetic-energy cutoff of 400 eV. Owing to the large periodic supercells and the absence of long-range crystalline order, Brillouin-zone sampling was restricted to the $\Gamma$ point. A time step of 1 fs was used and a periodic boundary condition was used throughout the AIMD simulations.

The models presented in this work describe the atomic structure of \tdg~coatings well. We conclude this from extensive calculations of atomic structure features of the models. These features, summarized below, are averaged over the 10 models and standard errors are given when appropriate. In Fig. \ref{fig:PDF}, the computed x-ray total PDF from models are shown and compared with measured x-ray grazing incidence pair distribution functions (GIPDF) for films of \tdg~(Ti cation ratio = 40\%, as-deposited) from Ref \cite{prasai-submitted}. We see a broad agreement between the two PDF, including the alignment in the positions of major peaks and minima. We observe slight discrepancy in peak heights at 3.20 \AA~and 4.4 \AA, which may be coming from the use of theoretical x-ray scattering factors (obtained using Hartree--Fock wave functions; Ref \cite{cromer1968x}), presence of Ar atoms in the films on which x-ray measurements were made \cite{prasai-submitted} and Fourier transform artifacts from the finite cell sizes. 

To identify the correlations giving rise to the major peaks in total PDF, we also compute the number-density-based partial PDF (see Equation 8 from Ref \cite{keen2001comparison} for definition). The first peak in total RDF at 1.78 \AA~comes from Ge-O and Ti-O peaks. The second peak in x-ray PDF at 3.20 \AA~comes from cation-cation correlation. The O--O peak in partial PDF have no noticeable foot-print in x-ray total PDF because of low scattering cross-sections. It is clear that the models contain no ``wrong'' bonds and no bonds below $\approx$1.7 \AA. 

Beyond the agreement with the experimental PDF, the models capture the average local structure and expected disorder of amorphous TiO$_2$-doped GeO$_2$. Ge is mostly four- and five-fold coordinated, Ti is mainly four-, five-, and six-fold coordinated, and O is primarily two- and three-fold coordinated, with minor under- and over-coordinated populations. The bond-length and bond-angle distributions are broad, as expected for amorphous oxide networks \cite{walker2015electronic,prasai2012properties}, and the models contain the expected mixture of corner-, edge-, and face-sharing polyhedra. Ti--Ti connections show the largest fraction of edge- and face-sharing motifs, consistent with earlier TiO$_2$-based amorphous networks in which highly connected Ti environments are associated with local strain and electronic localization \cite{landmann2012fingerprints,mora2020disorder}. These trends agree with previous structural reports on TiO$_2$-doped GeO$_2$ \cite{prasai-submitted,zhang2026network,jiang2025machine}; full coordination, bond-angle, and polyhedral-connectivity statistics are given in the Supplementary Information \cite{supp}.

\begin{figure}[t]
	\centering
		\includegraphics[width=\linewidth]{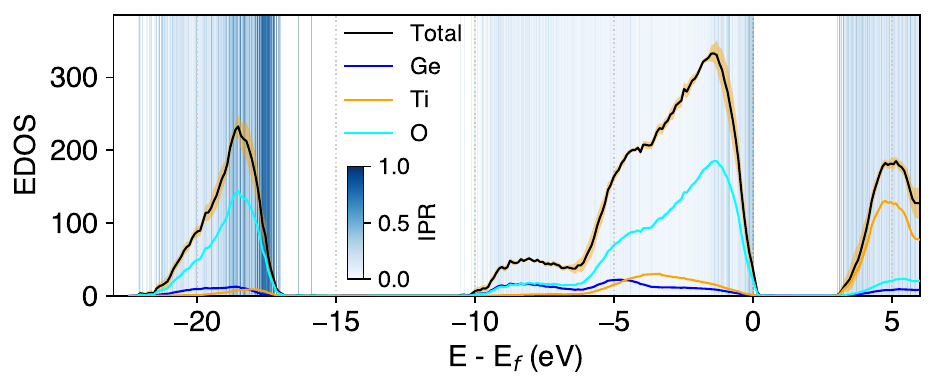}
		\caption{Electronic density of states (EDOS) and inverse participation ratio (IPR) for the stoichiometric amorphous TiGeO$_2$ model showing a clean gap and localized band-edge states. Site-projected EDOS, showing the O-dominated valence-band edge and cation-derived, primarily Ti-influenced, conduction-band edge.}
		\label{fig:EDOS}
\end{figure}

The electronic density of states (EDOS), defined as $D(E)=N^{-1}\sum_i^N\delta(E-E_i)$; where $N$ is size of the basis set and $E_i$ are Kohn--Sham eigenvalues, is plotted in Fig. \ref{fig:EDOS}. A measure of electronic localization, the inverse participation ratio (IPR), computed for the $n$th Kohn--Sham state $\psi_n$,

\vspace{-1.2em}
\begin{equation}
I_n=
\frac{
\sum_i \left|\langle \psi_n | \phi_i \rangle\right|^4
}{
\left(
\sum_i \left|\langle \psi_n | \phi_i \rangle\right|^2
\right)^2
},
\label{eq:ipr}
\end{equation}
\vspace{-1.2em}

\noindent where $\phi_i$ are local orbitals, is also plotted in Fig.~\ref{fig:EDOS}.

The calculated gap is approximately 3.73$\pm$0.07 eV. The experimentally determined value for thin films of 44\% \tdg~is $\sim$3.5 eV \cite{magnozzi2026exploring}. The IPR shows enhanced localization near the valence- and conduction-band edges, consistent with band-tail localization in an amorphous material. The projected EDOS shows that the upper valence band is dominated by O-derived states with strong $p$ character, while the lower conduction band is mainly cation-derived, with substantial Ti contribution and dominant $d$ character. Ge also contributes to the conduction manifold, but the lowest unoccupied states are strongly influenced by Ti. The central result is that the stoichiometric amorphous models retain a clean electronic gap. Although the structures contain distorted polyhedra, coordination fluctuations, and broad bond-angle distributions, these geometric irregularities do not produce deep midgap states. As the band gap is well above the 1064~nm photon energy of 1.17~eV, direct band-to-band absorption therefore cannot explain absorption at this wavelength in the stoichiometric amorphous network. The clean stoichiometric gap therefore provides a useful baseline: the experimentally relevant sub-gap absorption is unlikely to originate from generic amorphous disorder alone.

\input{table-main-v1}

\begin{figure}[h]
	\centering
		\includegraphics[width=\linewidth]{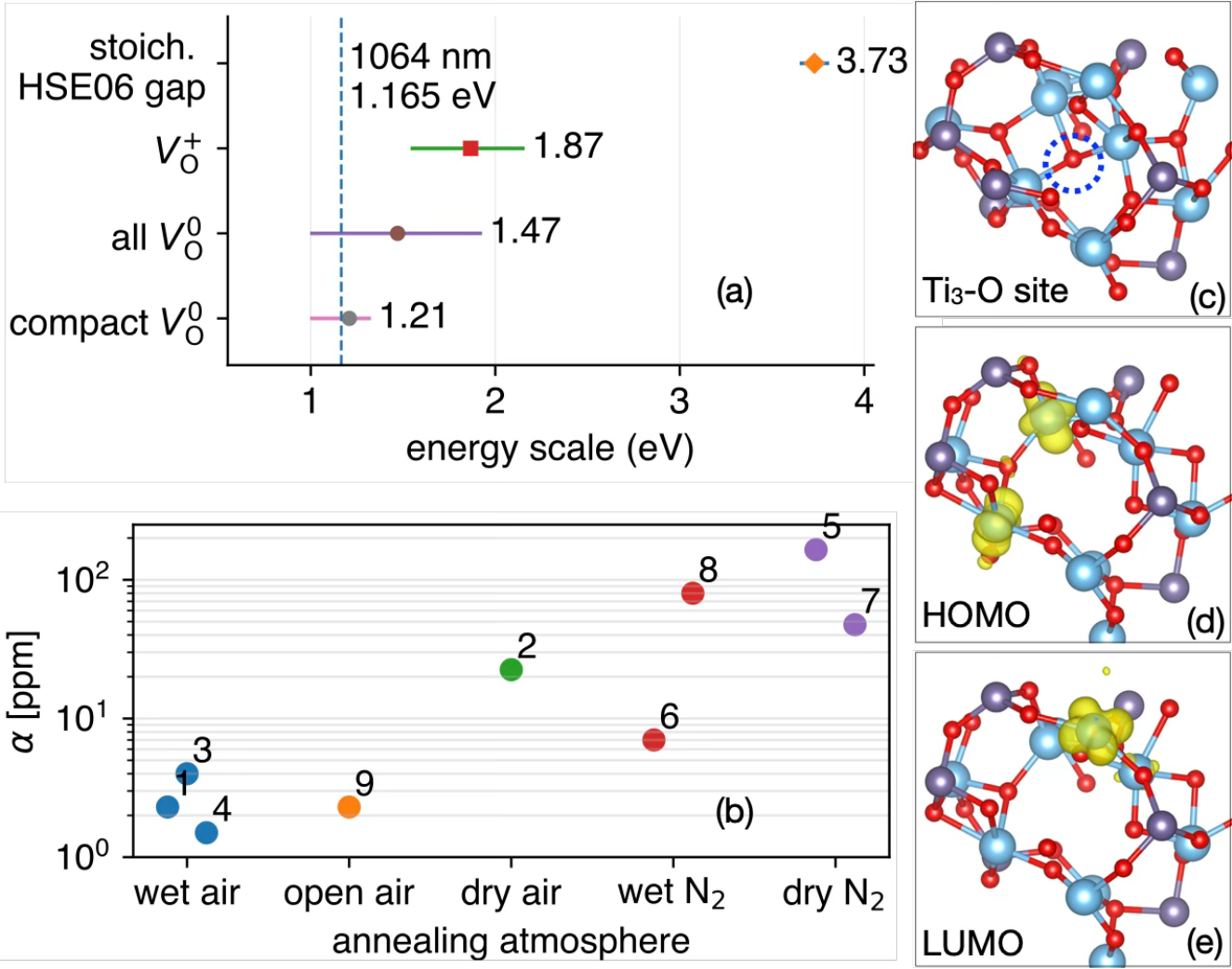}
		\caption{Oxygen-vacancy-induced near-infrared defect states in amorphous \tdg. (a) HSE06 energy scales comparing the stoichiometric amorphous gap with the 1064-nm photon energy and the defect-state separations produced by oxygen vacancies, (b) Atmosphere dependence of optical absorption in TiO$_2$-doped GeO$_2$ coating measured by photothermal common-path interferometry (PCI) at $\lambda=1064$ nm. Points show the measured absorption after sequential annealing treatments, with numbers indicating the treatment order, (c) Ti-rich precursor environment before oxygen removal. The selected O atom is coordinated to three nearby Ti atoms that form a compact local cluster, providing the structural motif that favors low-energy Ti-derived defect states, (d) the occupied state (HOMO) associated with the Ti-centered polaron pair, and (e) the lowest empty state (LUMO) as Ti-derived localized state in the same compact vacancy environment; their spin-conserving HSE06 separation is $\Delta\epsilon_{\rm HSE}=1.0065$ eV.}
		\label{fig:oxygen-defect}
\end{figure}

We next examined whether oxygen deficiency in Ti-rich local environments can introduce electronic states inside the otherwise clean gap of amorphous \tdg. Oxygen atoms were removed from Ti-rich environments in ten amorphous models, the defective structures were relaxed with PBE+\(U\) using several spin initializations, and static HSE06 calculations were performed on the relaxed structures. The key electronic quantity reported here is the smallest spin-conserving occupied-to-empty separation in the HSE06 HOMO-window spectrum, denoted $\Delta\epsilon_{\rm HSE}$. This quantity directly indicates whether localized defect states occur near the 1064-nm photon energy, $h\nu=1.165$ eV.

Neutral oxygen vacancies, $V_{\rm O}^{0}$, consistently produce localized Ti-derived sub-gap states. The neutral-vacancy HSE06 separations span 1.01--1.92 eV, with a mean of $1.49\pm0.04$ eV; see Fig.~\ref{fig:oxygen-defect} (a) and Table~\ref{tab:defect_summary_hse}. This broad range shows that oxygen removal alone does not determine the defect energy; the local Ti connectivity around the vacancy is decisive. The strongest near-1064-nm candidates are neutral vacancies in compact Ti-rich environments. In these branches, the vacancy is embedded in a compact Ti cluster and the HSE06 spin density localizes on Ti atoms adjacent to the missing O. Six compact neutral branches yield $\Delta\epsilon_{\rm HSE}=1.20\pm0.05$ eV, with a range of 1.01--1.32 eV, bracketing the 1064-nm photon energy. Representative cases include a mixed Ti-centered state with $\Delta\epsilon_{\rm HSE}=1.10$ eV, a Ti-polaron pair with $\Delta\epsilon_{\rm HSE}=1.20$ eV, and an antiferromagnetic Ti-centered pair with $\Delta\epsilon_{\rm HSE}=1.01$ eV. The latter is visualized as a charge-density-overlaid ball-and-stick model in Fig.~\ref{fig:oxygen-defect}, (c) to (e). Branches localized on more open Ti pairs, or vacancies with less compact Ti connectivity, generally shift to larger separations of $\sim1.5$--1.9 eV.

The singly positive vacancy, $V_{\rm O}^{+}$, forms a robust single-Ti-polaron state. In these calculations, the HSE06 spin density localizes on one Ti atom adjacent to the vacancy, with a moment close to $1~\mu_{\rm B}$. However, the corresponding HSE06 separations average $1.85\pm0.08$ eV, substantially above the 1064-nm photon energy. Thus, $V_{\rm O}^{+}$ demonstrates Ti$^{3+}$ polaron formation, but compact neutral vacancies are more plausible direct candidates for 1064-nm optical activity.

To assess whether the defect-state transitions are optically allowed, we computed HSE06 optical matrix elements, \(D^\alpha_{cv}=\langle \psi_c|\hat{D}_\alpha|\psi_v\rangle\), for each spin-conserving occupied-to-empty (\(v\rightarrow c\)) transition, with \(\alpha=x,y,z\). For the \(\Gamma\)-point amorphous supercells used here, we define the orientationally averaged transition strength as \(S_{\rm iso}(v,c)=\frac{1}{3}\sum_{\alpha=x,y,z}\left|D^\alpha_{cv}\right|^2\).  The optical matrix elements show that the allowedness of the defect-state transition is strongly configuration dependent. Compact Ti-rich neutral vacancies provide the relevant near-infrared energy scale, but the transition strength varies by orders of magnitude depending on whether the occupied and empty defect states remain strongly localized on separate Ti centers or mix within the same local Ti-rich cluster. The strongest near-1064-nm case is an antiferromagnetic Ti-polaron-pair branch, shown in Fig \ref{fig:oxygen-defect} (c) to (e), with \(\Delta\epsilon_{\rm HSE}=1.006~{\rm eV}\) and \(S_{\rm iso}=1.89\times10^{-2}~{\rm \AA}^{2}\), whereas other compact branches have much weaker but finite matrix elements (\(\sim 10^{-5}\)--\(10^{-3}~{\rm \AA}^{2}\)). Thus compact neutral oxygen vacancies are not merely energetically plausible; selected Ti-rich polaron-pair configurations also support optically allowed spin-conserving near-infrared transitions.

Absorption measurements provide an experimental constraint on this interpretation. Figure~\ref{fig:oxygen-defect} shows photothermal common-path interferometry (PCI) measurements at $\lambda=1064$~nm on single-layer ion-beam-sputtered TiO$_2$-doped GeO$_2$ coatings after atmosphere-dependent annealing. The absorption is strongly process dependent: wet-air or open-air treatments give low ppm-level absorption, whereas dry or inert annealing can increase the absorption by more than an order of magnitude. This behavior is difficult to reconcile with absorption from the ideal stoichiometric amorphous network alone, but is consistent with a reduction-sensitive population of defect centers. Together with the calculated clean gap of stoichiometric TiO$_2$-doped GeO$_2$ and the near-1064-nm Ti-derived states produced by neutral oxygen vacancies, the PCI trend supports oxygen-deficient, Ti-rich local environments as plausible contributors to the residual sub-gap absorption.

This interpretation is consistent with earlier work on titanium oxides and related Ti-doped wide-gap oxides. In crystalline TiO$_2$, excess carriers introduced by photoexcitation or reducing defects such as oxygen vacancies can self-trap as small polarons, producing localized charge densities coupled to local Ti--O lattice distortions~\cite{Chen2020JanusTiO2,Quirk2023TiO2PolaronRecombination}. A closely related precedent is Ti:sapphire, where first-principles calculations attributed residual near-infrared absorption to specific Ti ion-pair motifs, particularly line-contact antiferromagnetically coupled Ti$^{3+}$--Ti$^{3+}$ pairs and face-contact Ti$^{4+}$--Ti$^{3+}$ pairs~\cite{Gong2021TiSapphireIR}. These results support our interpretation of oxygen-deficient, Ti-rich motifs in amorphous TiO$_2$-doped GeO$_2$ as Ti-polaronic absorption centers, while the absence of comparable deep states in stoichiometric models indicates that such centers are associated with reduction and oxygen deficiency rather than ordinary amorphous disorder alone.

In conclusion, the residual 1064-nm absorption in TiO$_2$-doped GeO$_2$ does not appear to be an unavoidable consequence of the stoichiometric amorphous network. The stoichiometric models reproduce the measured structural disorder yet retain a clean HSE06 gap, showing that ordinary coordination disorder, strained polyhedra, and Ti-rich motifs alone are insufficient to generate deep near-infrared states. Oxygen deficiency changes this picture qualitatively. Neutral vacancies in compact Ti-rich environments stabilize localized Ti-derived polaron-pair or mixed-polaron states with separations that bracket the 1064-nm photon energy, and selected strong-overlap configurations have finite, and in some cases sizable, optical matrix elements. The strong dependence of the measured absorption on annealing atmosphere is consistent with this reduction-sensitive defect picture. These results identify oxygen-deficient compact Ti-rich motifs as microscopic targets for suppressing residual absorption in \tdg/\silica\ mirror coatings.

KP acknowledges support from the National Science Foundation (NSF), grant number 2513491, and the Gordon and Betty Moore Foundation (GBMF), grant GBMF6793.02. MMF acknowledges support from the NSF under grant numbers 2309289, 2309086, and 2513483, and from the GBMF under grant GBMF6793.02. KHL acknowledges support from the National Research Foundation of Korea (NRF), funded by the Korean government (MSIT), under grant No. RS-2024-00455482. This paper has LIGO document No. P2600260.

\bibliography{kbib}
\bibliographystyle{apsrev4-1}
\clearpage
\appendix
\include{supplemental}

\end{document}

%% file: table-main-v1.tex
\begin{table*}[t]
\caption{
HSE06 electronic signatures and optical allowedness of stoichiometric and oxygen-deficient amorphous \tdg.
For the stoichiometric network, the listed energy is the HSE06 band gap.
For vacancy defects, $\Delta\epsilon_{\rm HSE}$ is the smallest spin-conserving occupied-to-empty separation in the static HSE06 HOMO-window spectrum.
The column $N$ gives the number of spectra or models included in the average.
Ranges are given in parentheses.
The final column gives the orientationally averaged transition strength, $S_{\rm iso}$, for the lowest spin-conserving defect-state transition.
The strong-overlap $V_{\rm O}^{0}$ row is a subset of neutral-vacancy branches in which the occupied and empty defect states have substantial overlap within the same Ti-rich cluster.
One non-converged neutral branch is excluded from the statistical summary.}
\begin{ruledtabular}
\begin{tabular}{lcccc}
Class
& $N$
& HSE06 character
& Energy scale (eV)
& Allowedness, $S_{\rm iso}$ (\AA$^2$) \\
\hline
Stoichiometric
& 10
& clean gap
& $E_g=3.73\pm0.07$
& -- \\

All retained $V_{\rm O}^{0}$
& 29
& Ti-derived states
& $1.49\pm0.04$ $(1.01$--$1.92)$
& $6.5\times10^{-6}$--$2.1\times10^{-2}$ \\

Compact $V_{\rm O}^{0}$
& 6
& near-1064 mixed/pair states
& $1.20\pm0.05$ $(1.01$--$1.32)$
& up to $1.89\times10^{-2}$ \\

Strong-overlap $V_{\rm O}^{0}$
& 6
& mixed/FM/AFM Ti-cluster states
& $1.55\pm0.14$ $(1.01$--$1.92)$
& $6.9\times10^{-3}$--$2.1\times10^{-2}$ \\

$V_{\rm O}^{+}$
& 10
& single Ti polaron
& $1.85\pm0.08$ $(1.39$--$2.16)$
& $6.6\times10^{-5}$--$8.9\times10^{-3}$ \\
\end{tabular}
\end{ruledtabular}
\label{tab:defect_summary_hse}
\end{table*}

%% file: supplemental.tex
\section{Supplementary Information}

\subsection{Coordination Statistics}

\begin{table}[h]
  \centering
  \caption{Distribution of atomic coordination in the TiGeO$_2$ models. Percentages are averaged over ten independent models. Bond cutoff distance of 2.6 \AA\ is used.}
  \label{table:coord}

  \begin{tabular*}{\linewidth}{@{\extracolsep{\fill}}ccc@{}}
    \hline
    Atom & Coordination & Percentage (\%) \\
    \hline
    \multirow{5}{*}{Ge}
       & 2 & 0.14$\pm$0.13 \\
       & 3 & 0.42$\pm$0.20 \\
       & 4 & 80.56$\pm$1.08 \\
       & 5 & 18.47$\pm$1.16 \\
       & 6 & 0.42$\pm$0.20 \\
    \hline
    \multirow{4}{*}{Ti}
       & 4 & 14.38$\pm$1.36 \\
       & 5 & 55.83$\pm$2.51 \\
       & 6 & 27.92$\pm$2.42 \\
       & 7 & 1.88$\pm$0.46 \\
    \hline
    \multirow{4}{*}{O}
       & 1 & 0.42$\pm$0.10 \\
       & 2 & 70.63$\pm$0.76 \\
       & 3 & 28.46$\pm$0.79 \\
       & 4 & 0.50$\pm$0.11 \\
    \hline
  \end{tabular*}
\end{table}

\subsection{Polyhedral Connectivity}

\begin{table}[h]
\caption{Distribution of polyhedral connection motifs for Ge--Ge, Ge--Ti, and Ti--Ti nearest-neighbor pairs. Values are given as mean $\pm$ standard deviation. Bond cutoff distance of 2.6 \AA\ is used.}
\begin{ruledtabular}
\begin{tabular}{lccc}
Connection Type
& \begin{tabular}[c]{@{}c@{}}Corner-shared\\ (\%)\end{tabular}
& \begin{tabular}[c]{@{}c@{}}Edge-shared\\ (\%)\end{tabular}
& \begin{tabular}[c]{@{}c@{}}Face-shared\\ (\%)\end{tabular} \\
\hline
Ge--Ge
& $90.81\pm0.80$
& $9.09\pm0.82$
& $0.10\pm0.09$ \\

Ge--Ti
& $82.95\pm1.07$
& $16.80\pm1.09$
& $0.25\pm0.09$ \\

Ti--Ti
& $67.85\pm1.14$
& $31.03\pm1.14$
& $1.12\pm0.33$ \\
\end{tabular}
\end{ruledtabular}
\label{tab:polyhedral_connections}
\end{table}

\subsection{Bond Angle Distributions}

\begin{figure}[h]
	\centering
		\includegraphics[width=\linewidth]{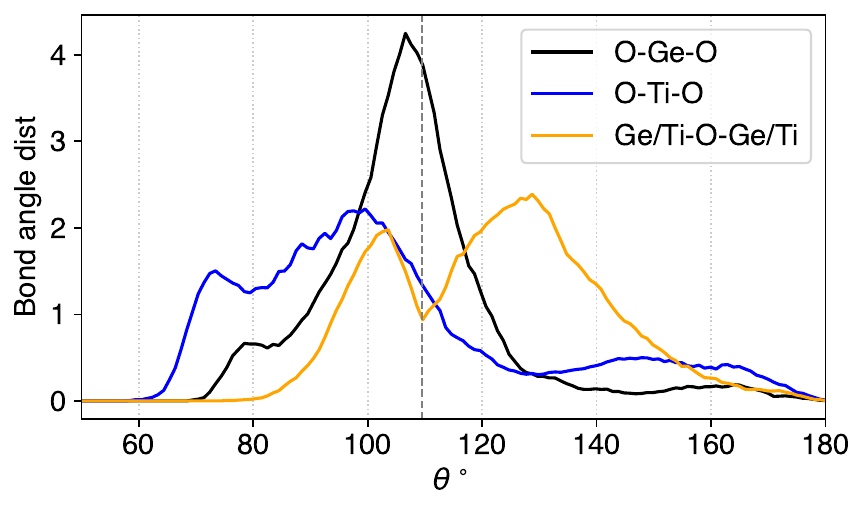}
		\caption{Bond-angle distributions in the amorphous TiO$_2$-doped GeO$_2$ models. The O--Ge--O and O--Ti--O distributions describe the local geometry of Ge- and Ti-centered oxygen polyhedra, while the Ge/Ti--O--Ge/Ti distribution describes inter-polyhedral connections through bridging oxygen atoms. The peak positions are similar to those reported in Ref \cite{walker2015electronic}  for a-\germania and Ref \cite{prasai2012properties} for a-\titania  -- but with broader distributions reflecting the structural disorder of the mixed amorphous network. Bond cutoff distance of 2.6 \AA\ is used.}
		\label{fig:BADF}
\end{figure}